\documentclass[aps,prl,reprint,twocolumn,twoside,showpacs,floatfix,letterpaper,superscriptaddress,bibnotes,footinbib]{revtex4-1}

\usepackage{graphicx}
\usepackage{amsmath}
\usepackage{amssymb}
\usepackage{hyperref}

\newcommand{\hMpc}{h^{-1}{\rm Mpc}}

\newcommand{\void}{\mathrm{v}}
\newcommand{\gal}{\mathrm{g}}
\newcommand{\matter}{\mathrm{m}}
\newcommand{\rs}{r_s}
\newcommand{\dc}{\delta_c}

\begin{document}

\title{Constraints on Cosmology and Gravity from the Dynamics of Voids}

\author{Nico Hamaus}
\email{hamaus@usm.lmu.de}
\affiliation{Universit\"ats-Sternwarte M\"unchen, Fakult\"at f\"ur Physik, Ludwig-Maximilians-Universit\"at M\"unchen,\\ Scheinerstr.~1, D-81679 M\"unchen, Germany}

\author{Alice Pisani}
\affiliation{\mbox{CPPM, CNRS/IN2P3, UMR 7346, Aix Marseille Universit\'e, 163 avenue de Luminy, F-13288 Marseille, France}}
\affiliation{Institut d'Astrophysique de Paris, UMR 7095, Sorbonne Universit\'es, UPMC Universit\'e Paris 06, \\ 98 bis boulevard Arago, F-75014 Paris, France}
\affiliation{CNRS, UMR 7095, Institut d'Astrophysique de Paris, 98 bis boulevard Arago, F-75014 Paris, France}

\author{P. M. Sutter}
\affiliation{Center for Cosmology and AstroParticle Physics, The Ohio State University, 191 West Woodruff Avenue, Columbus, Ohio 43210, USA}
\affiliation{INFN---National Institute for Nuclear Physics, via Valerio 2, I-34127 Trieste, Italy}
\affiliation{INAF---Osservatorio Astronomico di Trieste, via Tiepolo 11, I-34143 Trieste, Italy}

\author{Guilhem Lavaux}
\affiliation{Institut d'Astrophysique de Paris, UMR 7095, Sorbonne Universit\'es, UPMC Universit\'e Paris 06, \\ 98 bis boulevard Arago, F-75014 Paris, France}
\affiliation{CNRS, UMR 7095, Institut d'Astrophysique de Paris, 98 bis boulevard Arago, F-75014 Paris, France}

\author{St\'ephanie Escoffier}
\affiliation{\mbox{CPPM, CNRS/IN2P3, UMR 7346, Aix Marseille Universit\'e, 163 avenue de Luminy, F-13288 Marseille, France}}

\author{Benjamin D. Wandelt}
\affiliation{Institut d'Astrophysique de Paris, UMR 7095, Sorbonne Universit\'es, UPMC Universit\'e Paris 06, \\ 98 bis boulevard Arago, F-75014 Paris, France}
\affiliation{CNRS, UMR 7095, Institut d'Astrophysique de Paris, 98 bis boulevard Arago, F-75014 Paris, France}
\affiliation{Departments of Physics and Astronomy, University of Illinois at Urbana-Champaign, 1110 West Green Street, Urbana, Illinois 61801, USA}

\author{Jochen Weller}
\affiliation{Universit\"ats-Sternwarte M\"unchen, Fakult\"at f\"ur Physik, Ludwig-Maximilians-Universit\"at M\"unchen,\\ Scheinerstr.~1, D-81679 M\"unchen, Germany}
\affiliation{\mbox{Max Planck Institute for Extraterrestrial Physics, Giessenbachstr. 1, D-85748 Garching, Germany}}
\affiliation{Excellence Cluster Universe, Bolzmannstr. 2, D-85748 Garching, Germany}

\date{Received 25 February 2016; revised manuscript received 21 July 2016; published 25 August 2016}

\begin{abstract}
The Universe is mostly composed of large and relatively empty domains known as cosmic voids, whereas its matter content is predominantly distributed along their boundaries. The remaining material inside them, either dark or luminous matter, is attracted to these boundaries and causes voids to expand faster and to grow emptier over time. Using the distribution of galaxies centered on voids identified in the Sloan Digital Sky Survey and adopting minimal assumptions on the statistical motion of these galaxies, we constrain the average matter content $\Omega_\mathrm{m}=0.281\pm0.031$ in the Universe today, as well as the linear growth rate of structure $f/b=0.417\pm0.089$ at median redshift $\bar{z}=0.57$, where $b$ is the galaxy bias ($68\%$~C.L.). These values originate from a percent-level measurement of the anisotropic distortion in the void-galaxy cross-correlation function, $\varepsilon = 1.003\pm0.012$, and are robust to consistency tests with bootstraps of the data and simulated mock catalogs within an additional systematic uncertainty of half that size. They surpass (and are complementary to) existing constraints by unlocking cosmological information on smaller scales through an accurate model of nonlinear clustering and dynamics in void environments. As such, our analysis furnishes a powerful probe of deviations from Einstein's general relativity in the low-density regime which has largely remained untested so far. We find no evidence for such deviations in the data at hand.
\\ \\
DOI: \href{http://dx.doi.org/10.1103/PhysRevLett.117.091302}{10.1103/PhysRevLett.117.091302}
\end{abstract}


\maketitle

\textit{Introduction.}---After the epoch of recombination, the initially tiny Gaussian density perturbations in the early Universe have grown increasingly nonlinear under the influence of gravity, generating what is known as the \emph{cosmic web}. Because the gravitational force is attractive, structures with densities above the mean always contract in comoving coordinates, while underdense ones expand. The latter are referred to as \emph{cosmic voids} and have progressively occupied most of the available space in the Universe. Traditionally the formation of structure is viewed as hierarchical buildup of smaller dense clumps of matter into ever-larger objects. We take the dual perspective where structure formation is seen as the emptying out of void regions onto the walls, filaments, and clusters that surround them.

This void-centric point of view offers distinct advantages when probing the observed accelerated expansion of the Universe for two reasons: first, void dynamics are less nonlinear and, hence, more amenable to modeling than the high-density regime; second the accelerated expansion began at a density below the cosmic average. For this reason theories that attempt to explain the acceleration without introducing dark energy explicitly modify general relativity (GR) in the low-density regime. The effects of such modifications would therefore be most prominent in voids rather than in dense environments such as the solar system, galaxies, or clusters of galaxies.

While the dominant matter content of the Universe is invisible (dark), luminous tracers such as galaxies allow for the observation of the process of structure formation directly via their peculiar motions that follow the dynamics of voids. Although the individual velocity of galaxies cannot be determined in most cases, its line-of-sight component causes a Doppler shift in their spectrum, in addition to the Hubble redshift of each galaxy. This leads to a unique pattern of \emph{redshift-space distortions} (RSDs) in the distribution of galaxies around void centers, which allows for the inferring of their velocity flow statistically~\cite{Padilla2005,Paz2013,Micheletti2014}. The relation between galaxy density and velocity in voids can then be used to test the predictions of GR on cosmological scales~\cite{Hamaus2015}. So far most studies have focused on correlations between galaxies in this context, but in the dynamics of voids nonlinearities are less severe~\cite{Hamaus2014b,Hamaus2015}. As a consequence a large amount of smaller-scale information is unlocked for cosmological inference, resulting in a substantial decrease of statistical errors. 

Another type of distortion in the distribution of galaxies can be generated by the so-called \emph{Alcock-Paczy\'nski} (AP) effect~\cite{Alcock1979}. Galaxy surveys measure the redshifts $\delta z$ and angles $\delta\vartheta$ between any two galaxies on the sky, but these can only be converted to the correct comoving distances parallel ($r_\parallel$) and perpendicular ($r_\perp$) to the line of sight, if the expansion history and the geometry of the Universe is known,
\begin{equation}
r_\parallel = \frac{c}{H(z)}\delta z \;,\quad r_\perp = D_A(z)\,\delta\vartheta\;. \label{coords}
\end{equation}
The expansion history is described by the Hubble rate
\begin{equation}
 H(z) = H_0\sqrt{\Omega_\matter(1+z)^3+\Omega_k(1+z)^2+\Omega_\Lambda}\;, \label{H(z)}
\end{equation}
and the geometry by the angular diameter distance
\begin{equation}
 D_A(z) = \frac{c}{H_0\sqrt{-\Omega_k}}\sin\left(H_0\sqrt{-\Omega_k}\int_0^z \frac{1}{H(z')}\;\mathrm{d}z'\right)\;. \label{D_A(z)}
\end{equation}
These, in turn, depend on the Hubble constant $H_0$, the matter and energy content $\Omega_\matter$ and $\Omega_\Lambda$, and the curvature $\Omega_k$ of the Universe today. Therefore, a spherically symmetric structure may appear as an ellipsoid when incorrect cosmological parameters are assumed. The correct parameters can be obtained by demanding that the average shape of cosmic voids be spherically symmetric~\cite{Lavaux2012,Sutter2012b,Pisani2014,Sutter2014b,Hamaus2014c}, i.e., the ellipticity
\begin{equation}
\varepsilon := \frac{r_\parallel}{r_\perp} = \frac{D^\mathrm{true}_A(z)H^\mathrm{true}(z)}{D^\mathrm{fid}_A(z)H^\mathrm{fid}(z)}\;, \label{AP}
\end{equation}
be unity for the average distribution of galaxies around voids. In this case, $r_\parallel$ and $r_\perp$ refer to distances between galaxies and void centers with a total separation of $r = (r_\parallel^2 + r_\perp^2)^{1/2}$, and we distinguish between the unknown true and the assumed fiducial values of $D_A$ and $H$.

\textit{Model.}---In this Letter we apply these two concepts to voids identified in the distribution of galaxies observed with a redshift survey. Thereby, we closely follow the methodology presented in Ref.~\cite{Hamaus2015}, which has been extensively tested on simulated mock-galaxy catalogs. The starting point is the \emph{Gaussian streaming model}~\cite{Fisher1995}, providing the average distribution of galaxies around voids (in short: void stack) in redshift space via their cross-correlation function
\begin{equation}
 1+\xi_{\void\gal}(\mathbf{r}) = \int\frac{1+b\delta_{\void}(r)}{\sqrt{2\pi}\sigma_v} \exp\left[-\frac{\left(v_\parallel-v_\void(r)\frac{r_\parallel}{r}\right)^2}{2\sigma_v^2}\right]\mathrm{d}v_\parallel\;. \label{GSM}
\end{equation}
Here, $r$ and $v$ denote void-centric distances and velocities of galaxies in real space. Because distances are observed in redshift space, one has to take into account the contribution from peculiar motions,
\begin{equation}
 r_\parallel = \tilde{r}_\parallel - \frac{v_\parallel}{H(z)}(1+z)\;,
\end{equation}
where the tilde symbol indicates redshift space. Moreover, $b$ describes the linear bias parameter for galaxies and $\sigma_v$ their velocity dispersion. In simulations we have verified that the linear galaxy-bias assumption applies as long as the density fluctuations are moderate, i.e., $|\delta_{\void}(r)|\lesssim1$. The radial density profile of voids in real space can be parametrized with an empirical fitting function obtained from simulations, such as that given in Ref.~\cite{Hamaus2014b},
\begin{equation}
 \delta_{\void}(r) = \dc\,\frac{1-(r/\rs)^\alpha}{1+(r/r_\void)^\beta}\;, \label{HSW}
\end{equation}
with a central underdensity $\dc$, scale radius $\rs$, slopes $\alpha$ and $\beta$, and the effective void radius $r_\void$. The latter is not a free parameter, but determined via $r_\void=(3V_\void/4\pi)^{1/3}$, where $V_\void$ is the total volume of a void.
The velocity profile can be obtained via mass conservation~\cite{Peebles1980}. Up to linear order in density, it is given by
\begin{equation}
 v_\void(r) = - \frac{f(z)H(z)}{(1+z)r^2}\int_0^r \delta_{\void}(q)q^2 \,\mathrm{d}q \;, \label{v_lin}
\end{equation}
where $f(z)$ is the linear growth rate of density perturbations. Assuming GR and a flat $\Lambda$CDM cosmology it can be expressed as~\cite{Linder2005}
\begin{equation}
 f(z) \simeq \left(\frac{\Omega_\matter(1+z)^3}{\Omega_\matter(1+z)^3 + \Omega_\Lambda}\right)^{0.55} \;. \label{f(z)}
\end{equation}
Theories of modified gravity predict deviations from GR -- and thus Eq.~(\ref{f(z)}) -- to be most pronounced in unscreened low-density environments~\cite{Clifton2012}, potentially making voids a smoking gun for the detection of a fifth force. We have explicitly checked the range of validity for Eq.~(\ref{v_lin}) in the void environments we analyze using simulations~\cite{Hamaus2014b,Hamaus2015}. Note that the parameters $(f,b,\dc)$ are mutually degenerate in this model, but the combinations $f/b$ and $b\dc$ can be constrained independently.

\begin{figure*}
\centering
\resizebox{0.95\hsize}{!}{\includegraphics{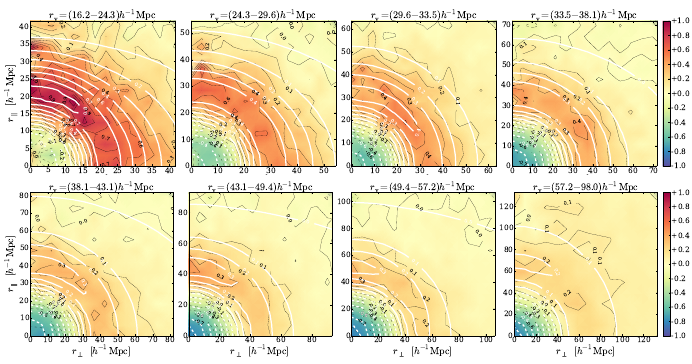}}
\caption{Void stacks from the SDSS-III DR11 CMASS galaxies at median redshift $\bar{z}=0.57$ in bins of increasing effective void radius $r_\void$. Void centers are at the origin and the statistical distribution of galaxies in void-centric distances along and perpendicular to the line of sight ($r_\parallel$, $r_\perp$) is color coded: red indicates more and blue, fewer galaxies than average. By construction the average is set to zero (yellow). Black solid (dashed) lines show positive (negative) contours of the data; white lines show the maximum-likelihood fit of the model. Because of the symmetry of the stacks, only one quadrant is shown. The enhanced ridge feature along $r_\parallel$ is caused by the coherent outflow of galaxies from the interior of voids. This allows us to infer the strength of gravity (growth rate $f/b$) when compared to directions perpendicular to the line of sight $r_\perp$.}
\label{stacks}
\end{figure*}

\textit{Data.}---Our results are shown in Fig.~\ref{stacks} for cosmic voids identified in the Sloan Digital Sky Survey (SDSS) DR11 at a median redshift $\bar{z}=0.57$~\cite{SM}. The different panels show void stacks of increasing effective void radius from left to right and top to bottom. Deviations from spherical symmetry are significant and clearly visible even by eye. These are due to RSDs caused by peculiar velocities in the statistical distribution of galaxies around voids. On large-enough scales most galaxies are attracted coherently by overdensities of the matter distribution and do not change directions, which leads to the characteristic compression of the ridge feature around the void centers along the line of sight. This squashing of overdensities in redshift space is known as the \emph{Kaiser} effect~\cite{Kaiser1987}. On smaller scales the velocity dispersion of galaxies becomes dominant over their coherent flow, causing an elongation of overdense structures along the line of sight that opposes the latter; this is commonly referred to as the \emph{finger-of-God} (FOG) effect. However, the scales considered in this analysis are still large, and the density fluctuations are small enough for the Kaiser effect to be the dominant one, as evident in Fig.~\ref{stacks}. It is also worth noticing the increase of central underdensities towards smaller voids, which is caused by finite-sampling effects when approaching the average galaxy separation of the sample. This effect does not, however, influence the anisotropic component of the void stacks, so it can be marginalized over via the free parameters in Eq.~(\ref{HSW}).

\begin{figure*}
\centering
\resizebox{0.95\hsize}{!}{\includegraphics{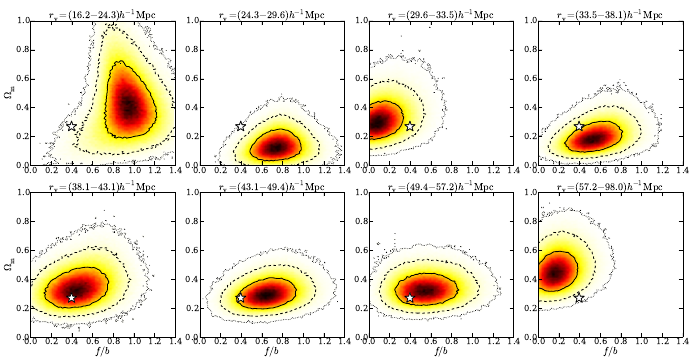}}
\caption{Constraints on matter density $\Omega_\matter$ and growth rate $f/b$ from each individual void stack of Fig.~\ref{stacks}. Solid, dashed, and dotted contour lines represent $68.3\%$, $95.5\%$, and $99.7\%$ credible regions, respectively. Stars indicate fiducial values of $\Omega_\matter=0.27$ and $f/b=0.40$.}
\label{pdfs}
\end{figure*}

\textit{Analysis.}---In order to compare our model from Eq.~(\ref{GSM}) with the observational data, we employ a Markov chain Monte Carlo (MCMC) technique~\cite{SM}. The best-fit solutions are shown as white contour levels in Fig.~\ref{stacks} and the posterior distributions in the $\Omega_\matter-f/b$ plane for the individual void stacks are shown in Fig.~\ref{pdfs}. In general a very reasonable agreement with our assumed fiducial cosmology is achieved, especially for intermediate-size voids within the range $30\hMpc \lesssim r_\void \lesssim 60\hMpc$. On smaller scales the effects of nonlinear RSDs (FoG) may cause systematic deviations that are not accounted for in our model~\cite{Hamaus2015}. On the other hand, our largest void stack necessarily exhibits the widest range of void sizes, as the void abundance drops exponentially in this regime. Therefore, both the RSD signal and the void profile get smeared over a wider range of scales, which can result in a biased fit. Nevertheless, the posteriors on $\Omega_\matter$ and $f/b$ are all consistent with each other across a wide range of scales, providing largely independent and competitive constraints to the existing literature.

\begin{figure}[!b]
\centering
\resizebox{0.9\hsize}{!}{\includegraphics{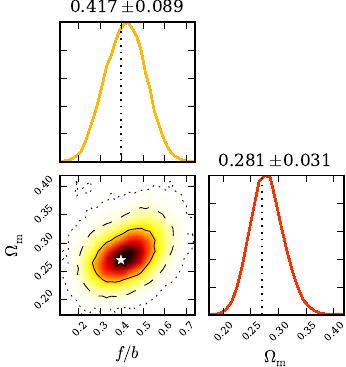}}
\caption{Joint constraints on matter density $\Omega_\matter$ and growth rate $f/b$ from all void stacks at median redshift $\bar{z}=0.57$ combined. Their mean and standard deviation is shown above the marginal distributions. The star and dotted lines indicate fiducial values of $\Omega_\matter=0.27$ and $f/b=0.40$.}
\label{pdfall}
\end{figure}

This is particularly the case when we choose to combine all the void stacks and infer the posterior parameter distribution jointly in a single MCMC chain that takes into account all the data at once. The resulting posterior distribution is presented in Fig.~\ref{pdfall}, including the marginal distributions for both $\Omega_\matter$ and $f/b$ individually. Our fiducial cosmology consistently falls inside the innermost confidence level of their joint posterior, and the standard deviation from the marginal distributions amounts to $\sim11\%$ for $\Omega_\matter$ and $\sim21\%$ for $f/b$, relative to their mean values. This implies $\varepsilon = 1.003\pm0.012$, a $\sim1\%$ precision on the AP parameter from Eq.~(\ref{AP}), which is nearly a factor of $4$ smaller than current state-of-the-art galaxy clustering constraints from RSDs (e.g., Ref.~\cite{Gil-Marin2016}), but obtained from a different regime of large-scale structure. We tested the robustness of our constraints using bootstraps of the data and mock catalogs and identify an additional systematic uncertainty of approximately $0.5\sigma$ caused by a residual dependence on the choice of our fiducial cosmology (see \cite{SM}). Moreover, so far we have neglected the large-scale regime of the void-galaxy cross-correlation function. It exhibits the baryon acoustic oscillation (BAO) feature, a relic clustering excess from the very early Universe. The latter provides a standard ruler and allows for the breaking of the degeneracy between $D_A(z)$ and $H(z)$ in Eq.~(\ref{AP}), resulting in even tighter cosmological constraints. The BAO feature in the clustering statistics of cosmic voids has recently been detected in the same data~\cite{Kitaura2016} (using a different void definition); it provides complementary information to the RSD analysis conducted in this Letter.

The consequences of modifications to GR are expected to be most striking in the low-density regime of the cosmic web~\cite{Clifton2012}. For example, voids extracted from simulations in $f(R)$ gravity exhibit significantly higher radial velocity flows compared to standard GR~\cite{Cai2015}. If present, this effect would be absorbed into our constraint on $f/b$ by biasing it high via Eq.~(\ref{v_lin}). We find no significant evidence for such a bias at the current level of precision.

\textit{Conclusions.}---Our analysis demonstrates that a substantial amount of unexplored cosmological information can be made available through the analysis of cosmic voids. Besides their dynamics studied in this Letter, voids also act as gravitational lenses~\cite{Melchior2014,Clampitt2014,Gruen2016}, exhibit rich clustering statistics~\cite{Hamaus2014a,ChuenChan2014,Clampitt2015} including the BAO feature~\cite{Kitaura2016}, and constrain cosmology through their abundance and shapes~\cite{Biswas2010,Pisani2015a}. These complementary cosmological observables break parameter degeneracies~\cite{Sahlen2016} and are promising probes of dark energy, GR~\cite{Cai2015,Zivick2015,Barreira2015}, or the impact of massive neutrinos~\cite{Massara2015} on cosmological scales. Different void finders most likely yield various trade-offs between the strength of the sought-after signal and the ability to model it, so the optimal void definition will depend on the specific application. We leave further investigations along these lines to future work.
\newline

\begin{acknowledgments}
We thank D. Paz, A. Hawken, B. Hoyle, and M.-C. Cousinou for discussions. Computations were performed on the HORIZON cluster at IAP. This work was supported by the ILP LABEX (Grant No. ANR-10-LABX-63), French state funds managed by the ANR within the ``Investissements d'Avenir'' program (Grant No. ANR-11-IDEX-0004-02), and NSF Grant No. AST 09-08693 ARRA. N.~H. and J.~W. acknowledge support from the DFG cluster of excellence ``Origin and Structure of the Universe''. A.~P. acknowledges financial support from OMEGA Grant No. ANR-11-JS56-003-01 and the support of the OCEVU LABEX (Grant No. ANR-11-LABX-0060) and the A*MIDEX project (Grant No. ANR-11-IDEX-0001-02) funded by the ``Investissements d'Avenir''  French government program managed by the ANR. P.~M.~S. is supported by the INFN IS PD51 ``Indark''. B.~D.~W. is supported by a senior Excellence Chair by the Agence Nationale de Recherche (Grant No. ANR-10-CEXC-004-01) and a Chaire Internationale at the Universit\'e Pierre et Marie Curie. J.~W. also acknowledges support from the Trans-Regional Collaborative Research Center TRR 33 ``The Dark Universe'' of the Deutsche Forschungsgemeinschaft (DFG).

Funding for SDSS-III has been provided by the Alfred P. Sloan Foundation and its participating institutions, the National Science Foundation, and the U.S. Department of Energy Office of Science. The SDSS-III web site is \url{www.sdss3.org}. SDSS-III is managed by the Astrophysical Research Consortium for the Participating Institutions of the SDSS-III Collaboration including the University of Arizona, the Brazilian Participation Group, Brookhaven National Laboratory, Carnegie Mellon University, University of Florida, the French Participation Group, the German Participation Group, Harvard University, the Instituto de Astrofisica de Canarias, the Michigan State/Notre Dame/JINA Participation Group, Johns Hopkins University, Lawrence Berkeley National Laboratory, Max Planck Institute for Astrophysics, Max Planck Institute for Extraterrestrial Physics, New Mexico State University, New York University, Ohio State University, Pennsylvania State University, University of Portsmouth, Princeton University, the Spanish Participation Group, University of Tokyo, University of Utah, Vanderbilt University, University of Virginia, University of Washington, and Yale University. \newline
\end{acknowledgments}

\bibliography{ms.bib}

\begin{thebibliography}{48}
\expandafter\ifx\csname natexlab\endcsname\relax\def\natexlab#1{#1}\fi
\expandafter\ifx\csname bibnamefont\endcsname\relax
  \def\bibnamefont#1{#1}\fi
\expandafter\ifx\csname bibfnamefont\endcsname\relax
  \def\bibfnamefont#1{#1}\fi
\expandafter\ifx\csname citenamefont\endcsname\relax
  \def\citenamefont#1{#1}\fi
\expandafter\ifx\csname url\endcsname\relax
  \def\url#1{\texttt{#1}}\fi
\expandafter\ifx\csname urlprefix\endcsname\relax\def\urlprefix{URL }\fi
\providecommand{\bibinfo}[2]{#2}
\providecommand{\eprint}[2][]{\url{#2}}

\bibitem[{\citenamefont{{Padilla} et~al.}(2005)\citenamefont{{Padilla},
  {Ceccarelli}, and {Lambas}}}]{Padilla2005}
\bibinfo{author}{\bibfnamefont{N.~D.} \bibnamefont{{Padilla}}},
  \bibinfo{author}{\bibfnamefont{L.}~\bibnamefont{{Ceccarelli}}},
  \bibnamefont{and} \bibinfo{author}{\bibfnamefont{D.~G.}
  \bibnamefont{{Lambas}}}, \bibinfo{journal}{\mnras}
  \textbf{\bibinfo{volume}{363}}, \bibinfo{pages}{977} (\bibinfo{year}{2005}),
  \eprint{astro-ph/0508297}.

\bibitem[{\citenamefont{{Paz} et~al.}(2013)\citenamefont{{Paz}, {Lares},
  {Ceccarelli}, {Padilla}, and {Lambas}}}]{Paz2013}
\bibinfo{author}{\bibfnamefont{D.}~\bibnamefont{{Paz}}},
  \bibinfo{author}{\bibfnamefont{M.}~\bibnamefont{{Lares}}},
  \bibinfo{author}{\bibfnamefont{L.}~\bibnamefont{{Ceccarelli}}},
  \bibinfo{author}{\bibfnamefont{N.}~\bibnamefont{{Padilla}}},
  \bibnamefont{and} \bibinfo{author}{\bibfnamefont{D.~G.}
  \bibnamefont{{Lambas}}}, \bibinfo{journal}{\mnras}
  \textbf{\bibinfo{volume}{436}}, \bibinfo{pages}{3480} (\bibinfo{year}{2013}),
  \eprint{1306.5799}.

\bibitem[{\citenamefont{{Micheletti} et~al.}(2014)\citenamefont{{Micheletti},
  {Iovino}, {Hawken}, {Granett}, {Bolzonella}, {Cappi}, {Guzzo}, {Abbas},
  {Adami}, {Arnouts} et~al.}}]{Micheletti2014}
\bibinfo{author}{\bibfnamefont{D.}~\bibnamefont{{Micheletti}}},
  \bibinfo{author}{\bibfnamefont{A.}~\bibnamefont{{Iovino}}},
  \bibinfo{author}{\bibfnamefont{A.~J.} \bibnamefont{{Hawken}}},
  \bibinfo{author}{\bibfnamefont{B.~R.} \bibnamefont{{Granett}}},
  \bibinfo{author}{\bibfnamefont{M.}~\bibnamefont{{Bolzonella}}},
  \bibinfo{author}{\bibfnamefont{A.}~\bibnamefont{{Cappi}}},
  \bibinfo{author}{\bibfnamefont{L.}~\bibnamefont{{Guzzo}}},
  \bibinfo{author}{\bibfnamefont{U.}~\bibnamefont{{Abbas}}},
  \bibinfo{author}{\bibfnamefont{C.}~\bibnamefont{{Adami}}},
  \bibinfo{author}{\bibfnamefont{S.}~\bibnamefont{{Arnouts}}},
  \bibnamefont{et~al.}, \bibinfo{journal}{\aap} \textbf{\bibinfo{volume}{570}},
  \bibinfo{eid}{A106} (\bibinfo{year}{2014}), \eprint{1407.2969}.

\bibitem[{\citenamefont{{Hamaus} et~al.}(2015)\citenamefont{{Hamaus}, {Sutter},
  {Lavaux}, and {Wandelt}}}]{Hamaus2015}
\bibinfo{author}{\bibfnamefont{N.}~\bibnamefont{{Hamaus}}},
  \bibinfo{author}{\bibfnamefont{P.~M.} \bibnamefont{{Sutter}}},
  \bibinfo{author}{\bibfnamefont{G.}~\bibnamefont{{Lavaux}}}, \bibnamefont{and}
  \bibinfo{author}{\bibfnamefont{B.~D.} \bibnamefont{{Wandelt}}},
  \bibinfo{journal}{\jcap} \textbf{\bibinfo{volume}{11}}, \bibinfo{eid}{036}
  (\bibinfo{year}{2015}), \eprint{1507.04363}.

\bibitem[{\citenamefont{{Hamaus}
  et~al.}(2014{\natexlab{a}})\citenamefont{{Hamaus}, {Sutter}, and
  {Wandelt}}}]{Hamaus2014b}
\bibinfo{author}{\bibfnamefont{N.}~\bibnamefont{{Hamaus}}},
  \bibinfo{author}{\bibfnamefont{P.~M.} \bibnamefont{{Sutter}}},
  \bibnamefont{and} \bibinfo{author}{\bibfnamefont{B.~D.}
  \bibnamefont{{Wandelt}}}, \bibinfo{journal}{\prl}
  \textbf{\bibinfo{volume}{112}}, \bibinfo{eid}{251302}
  (\bibinfo{year}{2014}{\natexlab{a}}), \eprint{1403.5499}.

\bibitem[{\citenamefont{{Alcock} and {Paczynski}}(1979)}]{Alcock1979}
\bibinfo{author}{\bibfnamefont{C.}~\bibnamefont{{Alcock}}} \bibnamefont{and}
  \bibinfo{author}{\bibfnamefont{B.}~\bibnamefont{{Paczynski}}},
  \bibinfo{journal}{\nat} \textbf{\bibinfo{volume}{281}}, \bibinfo{pages}{358}
  (\bibinfo{year}{1979}).

\bibitem[{\citenamefont{{Lavaux} and {Wandelt}}(2012)}]{Lavaux2012}
\bibinfo{author}{\bibfnamefont{G.}~\bibnamefont{{Lavaux}}} \bibnamefont{and}
  \bibinfo{author}{\bibfnamefont{B.~D.} \bibnamefont{{Wandelt}}},
  \bibinfo{journal}{\apj} \textbf{\bibinfo{volume}{754}}, \bibinfo{eid}{109}
  (\bibinfo{year}{2012}), \eprint{1110.0345}.

\bibitem[{\citenamefont{{Sutter}
  et~al.}(2012{\natexlab{a}})\citenamefont{{Sutter}, {Lavaux}, {Wandelt}, and
  {Weinberg}}}]{Sutter2012b}
\bibinfo{author}{\bibfnamefont{P.~M.} \bibnamefont{{Sutter}}},
  \bibinfo{author}{\bibfnamefont{G.}~\bibnamefont{{Lavaux}}},
  \bibinfo{author}{\bibfnamefont{B.~D.} \bibnamefont{{Wandelt}}},
  \bibnamefont{and} \bibinfo{author}{\bibfnamefont{D.~H.}
  \bibnamefont{{Weinberg}}}, \bibinfo{journal}{\apj}
  \textbf{\bibinfo{volume}{761}}, \bibinfo{eid}{187}
  (\bibinfo{year}{2012}{\natexlab{a}}), \eprint{1208.1058}.

\bibitem[{\citenamefont{{Pisani} et~al.}(2014)\citenamefont{{Pisani}, {Lavaux},
  {Sutter}, and {Wandelt}}}]{Pisani2014}
\bibinfo{author}{\bibfnamefont{A.}~\bibnamefont{{Pisani}}},
  \bibinfo{author}{\bibfnamefont{G.}~\bibnamefont{{Lavaux}}},
  \bibinfo{author}{\bibfnamefont{P.~M.} \bibnamefont{{Sutter}}},
  \bibnamefont{and} \bibinfo{author}{\bibfnamefont{B.~D.}
  \bibnamefont{{Wandelt}}}, \bibinfo{journal}{\mnras}
  \textbf{\bibinfo{volume}{443}}, \bibinfo{pages}{3238} (\bibinfo{year}{2014}),
  \eprint{1306.3052}.

\bibitem[{\citenamefont{{Sutter}
  et~al.}(2014{\natexlab{a}})\citenamefont{{Sutter}, {Pisani}, {Wandelt}, and
  {Weinberg}}}]{Sutter2014b}
\bibinfo{author}{\bibfnamefont{P.~M.} \bibnamefont{{Sutter}}},
  \bibinfo{author}{\bibfnamefont{A.}~\bibnamefont{{Pisani}}},
  \bibinfo{author}{\bibfnamefont{B.~D.} \bibnamefont{{Wandelt}}},
  \bibnamefont{and} \bibinfo{author}{\bibfnamefont{D.~H.}
  \bibnamefont{{Weinberg}}}, \bibinfo{journal}{\mnras}
  \textbf{\bibinfo{volume}{443}}, \bibinfo{pages}{2983}
  (\bibinfo{year}{2014}{\natexlab{a}}), \eprint{1404.5618}.

\bibitem[{\citenamefont{{Hamaus}
  et~al.}(2014{\natexlab{b}})\citenamefont{{Hamaus}, {Sutter}, {Lavaux}, and
  {Wandelt}}}]{Hamaus2014c}
\bibinfo{author}{\bibfnamefont{N.}~\bibnamefont{{Hamaus}}},
  \bibinfo{author}{\bibfnamefont{P.~M.} \bibnamefont{{Sutter}}},
  \bibinfo{author}{\bibfnamefont{G.}~\bibnamefont{{Lavaux}}}, \bibnamefont{and}
  \bibinfo{author}{\bibfnamefont{B.~D.} \bibnamefont{{Wandelt}}},
  \bibinfo{journal}{\jcap} \textbf{\bibinfo{volume}{12}}, \bibinfo{eid}{013}
  (\bibinfo{year}{2014}{\natexlab{b}}), \eprint{1409.3580}.

\bibitem[{\citenamefont{{Fisher}}(1995)}]{Fisher1995}
\bibinfo{author}{\bibfnamefont{K.~B.} \bibnamefont{{Fisher}}},
  \bibinfo{journal}{\apj} \textbf{\bibinfo{volume}{448}}, \bibinfo{pages}{494}
  (\bibinfo{year}{1995}), \eprint{astro-ph/9412081}.

\bibitem[{\citenamefont{{Peebles}}(1980)}]{Peebles1980}
\bibinfo{author}{\bibfnamefont{P.~J.~E.} \bibnamefont{{Peebles}}},
  \emph{\bibinfo{title}{{The Large-Scale Structure of the Universe}}}
  (\bibinfo{publisher}{Princeton University Press, Princeton, NJ},
  \bibinfo{year}{1980}).

\bibitem[{\citenamefont{{Linder}}(2005)}]{Linder2005}
\bibinfo{author}{\bibfnamefont{E.~V.} \bibnamefont{{Linder}}},
  \bibinfo{journal}{\prd} \textbf{\bibinfo{volume}{72}}, \bibinfo{eid}{043529}
  (\bibinfo{year}{2005}), \eprint{astro-ph/0507263}.

\bibitem[{\citenamefont{{Clifton} et~al.}(2012)\citenamefont{{Clifton},
  {Ferreira}, {Padilla}, and {Skordis}}}]{Clifton2012}
\bibinfo{author}{\bibfnamefont{T.}~\bibnamefont{{Clifton}}},
  \bibinfo{author}{\bibfnamefont{P.~G.} \bibnamefont{{Ferreira}}},
  \bibinfo{author}{\bibfnamefont{A.}~\bibnamefont{{Padilla}}},
  \bibnamefont{and}
  \bibinfo{author}{\bibfnamefont{C.}~\bibnamefont{{Skordis}}},
  \bibinfo{journal}{\physrep} \textbf{\bibinfo{volume}{513}},
  \bibinfo{pages}{1} (\bibinfo{year}{2012}), \eprint{1106.2476}.

\bibitem[{SM()}]{SM}
\bibinfo{note}{See Supplemental Material below for details on the data
  analysis.}

\bibitem[{\citenamefont{{Kaiser}}(1987)}]{Kaiser1987}
\bibinfo{author}{\bibfnamefont{N.}~\bibnamefont{{Kaiser}}},
  \bibinfo{journal}{\mnras} \textbf{\bibinfo{volume}{227}}, \bibinfo{pages}{1}
  (\bibinfo{year}{1987}).

\bibitem[{\citenamefont{{Gil-Mar{\'{\i}}n}
  et~al.}(2016)\citenamefont{{Gil-Mar{\'{\i}}n}, {Percival}, {Brownstein},
  {Chuang}, {Grieb}, {Ho}, {Kitaura}, {Maraston}, {Prada},
  {Rodr{\'{\i}}guez-Torres} et~al.}}]{Gil-Marin2016}
\bibinfo{author}{\bibfnamefont{H.}~\bibnamefont{{Gil-Mar{\'{\i}}n}}},
  \bibinfo{author}{\bibfnamefont{W.~J.} \bibnamefont{{Percival}}},
  \bibinfo{author}{\bibfnamefont{J.~R.} \bibnamefont{{Brownstein}}},
  \bibinfo{author}{\bibfnamefont{C.-H.} \bibnamefont{{Chuang}}},
  \bibinfo{author}{\bibfnamefont{J.~N.} \bibnamefont{{Grieb}}},
  \bibinfo{author}{\bibfnamefont{S.}~\bibnamefont{{Ho}}},
  \bibinfo{author}{\bibfnamefont{F.-S.} \bibnamefont{{Kitaura}}},
  \bibinfo{author}{\bibfnamefont{C.}~\bibnamefont{{Maraston}}},
  \bibinfo{author}{\bibfnamefont{F.}~\bibnamefont{{Prada}}},
  \bibinfo{author}{\bibfnamefont{S.}~\bibnamefont{{Rodr{\'{\i}}guez-Torres}}},
  \bibnamefont{et~al.}, \bibinfo{journal}{\mnras}
  \textbf{\bibinfo{volume}{460}}, \bibinfo{pages}{4188} (\bibinfo{year}{2016}),
  \eprint{1509.06386}.

\bibitem[{\citenamefont{{Kitaura} et~al.}(2016)\citenamefont{{Kitaura},
  {Chuang}, {Liang}, {Zhao}, {Tao}, {Rodr{\'{\i}}guez-Torres}, {Eisenstein},
  {Gil-Mar{\'{\i}}n}, {Kneib}, {McBride} et~al.}}]{Kitaura2016}
\bibinfo{author}{\bibfnamefont{F.-S.} \bibnamefont{{Kitaura}}},
  \bibinfo{author}{\bibfnamefont{C.-H.} \bibnamefont{{Chuang}}},
  \bibinfo{author}{\bibfnamefont{Y.}~\bibnamefont{{Liang}}},
  \bibinfo{author}{\bibfnamefont{C.}~\bibnamefont{{Zhao}}},
  \bibinfo{author}{\bibfnamefont{C.}~\bibnamefont{{Tao}}},
  \bibinfo{author}{\bibfnamefont{S.}~\bibnamefont{{Rodr{\'{\i}}guez-Torres}}},
  \bibinfo{author}{\bibfnamefont{D.~J.} \bibnamefont{{Eisenstein}}},
  \bibinfo{author}{\bibfnamefont{H.}~\bibnamefont{{Gil-Mar{\'{\i}}n}}},
  \bibinfo{author}{\bibfnamefont{J.-P.} \bibnamefont{{Kneib}}},
  \bibinfo{author}{\bibfnamefont{C.}~\bibnamefont{{McBride}}},
  \bibnamefont{et~al.}, \bibinfo{journal}{\prl} \textbf{\bibinfo{volume}{116}},
  \bibinfo{eid}{171301} (\bibinfo{year}{2016}), \eprint{1511.04405}.

\bibitem[{\citenamefont{{Cai} et~al.}(2015)\citenamefont{{Cai}, {Padilla}, and
  {Li}}}]{Cai2015}
\bibinfo{author}{\bibfnamefont{Y.-C.} \bibnamefont{{Cai}}},
  \bibinfo{author}{\bibfnamefont{N.}~\bibnamefont{{Padilla}}},
  \bibnamefont{and} \bibinfo{author}{\bibfnamefont{B.}~\bibnamefont{{Li}}},
  \bibinfo{journal}{\mnras} \textbf{\bibinfo{volume}{451}},
  \bibinfo{pages}{1036} (\bibinfo{year}{2015}), \eprint{1410.1510}.

\bibitem[{\citenamefont{{Melchior} et~al.}(2014)\citenamefont{{Melchior},
  {Sutter}, {Sheldon}, {Krause}, and {Wandelt}}}]{Melchior2014}
\bibinfo{author}{\bibfnamefont{P.}~\bibnamefont{{Melchior}}},
  \bibinfo{author}{\bibfnamefont{P.~M.} \bibnamefont{{Sutter}}},
  \bibinfo{author}{\bibfnamefont{E.~S.} \bibnamefont{{Sheldon}}},
  \bibinfo{author}{\bibfnamefont{E.}~\bibnamefont{{Krause}}}, \bibnamefont{and}
  \bibinfo{author}{\bibfnamefont{B.~D.} \bibnamefont{{Wandelt}}},
  \bibinfo{journal}{\mnras} \textbf{\bibinfo{volume}{440}},
  \bibinfo{pages}{2922} (\bibinfo{year}{2014}), \eprint{1309.2045}.

\bibitem[{\citenamefont{{Clampitt} and {Jain}}(2015)}]{Clampitt2014}
\bibinfo{author}{\bibfnamefont{J.}~\bibnamefont{{Clampitt}}} \bibnamefont{and}
  \bibinfo{author}{\bibfnamefont{B.}~\bibnamefont{{Jain}}},
  \bibinfo{journal}{\mnras} \textbf{\bibinfo{volume}{454}},
  \bibinfo{pages}{3357} (\bibinfo{year}{2015}), \eprint{1404.1834}.

\bibitem[{\citenamefont{{Gruen} et~al.}(2016)\citenamefont{{Gruen},
  {Friedrich}, {Amara}, {Bacon}, {Bonnett}, {Hartley}, {Jain}, {Jarvis},
  {Kacprzak}, {Krause} et~al.}}]{Gruen2016}
\bibinfo{author}{\bibfnamefont{D.}~\bibnamefont{{Gruen}}},
  \bibinfo{author}{\bibfnamefont{O.}~\bibnamefont{{Friedrich}}},
  \bibinfo{author}{\bibfnamefont{A.}~\bibnamefont{{Amara}}},
  \bibinfo{author}{\bibfnamefont{D.}~\bibnamefont{{Bacon}}},
  \bibinfo{author}{\bibfnamefont{C.}~\bibnamefont{{Bonnett}}},
  \bibinfo{author}{\bibfnamefont{W.}~\bibnamefont{{Hartley}}},
  \bibinfo{author}{\bibfnamefont{B.}~\bibnamefont{{Jain}}},
  \bibinfo{author}{\bibfnamefont{M.}~\bibnamefont{{Jarvis}}},
  \bibinfo{author}{\bibfnamefont{T.}~\bibnamefont{{Kacprzak}}},
  \bibinfo{author}{\bibfnamefont{E.}~\bibnamefont{{Krause}}},
  \bibnamefont{et~al.}, \bibinfo{journal}{\mnras}
  \textbf{\bibinfo{volume}{455}}, \bibinfo{pages}{3367} (\bibinfo{year}{2016}),
  \eprint{1507.05090}.

\bibitem[{\citenamefont{{Hamaus}
  et~al.}(2014{\natexlab{c}})\citenamefont{{Hamaus}, {Wandelt}, {Sutter},
  {Lavaux}, and {Warren}}}]{Hamaus2014a}
\bibinfo{author}{\bibfnamefont{N.}~\bibnamefont{{Hamaus}}},
  \bibinfo{author}{\bibfnamefont{B.~D.} \bibnamefont{{Wandelt}}},
  \bibinfo{author}{\bibfnamefont{P.~M.} \bibnamefont{{Sutter}}},
  \bibinfo{author}{\bibfnamefont{G.}~\bibnamefont{{Lavaux}}}, \bibnamefont{and}
  \bibinfo{author}{\bibfnamefont{M.~S.} \bibnamefont{{Warren}}},
  \bibinfo{journal}{\prl} \textbf{\bibinfo{volume}{112}}, \bibinfo{eid}{041304}
  (\bibinfo{year}{2014}{\natexlab{c}}), \eprint{1307.2571}.

\bibitem[{\citenamefont{{Chan} et~al.}(2014)\citenamefont{{Chan}, {Hamaus}, and
  {Desjacques}}}]{ChuenChan2014}
\bibinfo{author}{\bibfnamefont{K.~C.} \bibnamefont{{Chan}}},
  \bibinfo{author}{\bibfnamefont{N.}~\bibnamefont{{Hamaus}}}, \bibnamefont{and}
  \bibinfo{author}{\bibfnamefont{V.}~\bibnamefont{{Desjacques}}},
  \bibinfo{journal}{\prd} \textbf{\bibinfo{volume}{90}}, \bibinfo{eid}{103521}
  (\bibinfo{year}{2014}).

\bibitem[{\citenamefont{{Clampitt} et~al.}(2016)\citenamefont{{Clampitt},
  {Jain}, and {S{\'a}nchez}}}]{Clampitt2015}
\bibinfo{author}{\bibfnamefont{J.}~\bibnamefont{{Clampitt}}},
  \bibinfo{author}{\bibfnamefont{B.}~\bibnamefont{{Jain}}}, \bibnamefont{and}
  \bibinfo{author}{\bibfnamefont{C.}~\bibnamefont{{S{\'a}nchez}}},
  \bibinfo{journal}{\mnras} \textbf{\bibinfo{volume}{456}},
  \bibinfo{pages}{4425} (\bibinfo{year}{2016}), \eprint{1507.08031}.

\bibitem[{\citenamefont{{Biswas} et~al.}(2010)\citenamefont{{Biswas},
  {Alizadeh}, and {Wandelt}}}]{Biswas2010}
\bibinfo{author}{\bibfnamefont{R.}~\bibnamefont{{Biswas}}},
  \bibinfo{author}{\bibfnamefont{E.}~\bibnamefont{{Alizadeh}}},
  \bibnamefont{and} \bibinfo{author}{\bibfnamefont{B.~D.}
  \bibnamefont{{Wandelt}}}, \bibinfo{journal}{\prd}
  \textbf{\bibinfo{volume}{82}}, \bibinfo{eid}{023002} (\bibinfo{year}{2010}),
  \eprint{1002.0014}.

\bibitem[{\citenamefont{{Pisani}
  et~al.}(2015{\natexlab{a}})\citenamefont{{Pisani}, {Sutter}, {Hamaus},
  {Alizadeh}, {Biswas}, {Wandelt}, and {Hirata}}}]{Pisani2015a}
\bibinfo{author}{\bibfnamefont{A.}~\bibnamefont{{Pisani}}},
  \bibinfo{author}{\bibfnamefont{P.~M.} \bibnamefont{{Sutter}}},
  \bibinfo{author}{\bibfnamefont{N.}~\bibnamefont{{Hamaus}}},
  \bibinfo{author}{\bibfnamefont{E.}~\bibnamefont{{Alizadeh}}},
  \bibinfo{author}{\bibfnamefont{R.}~\bibnamefont{{Biswas}}},
  \bibinfo{author}{\bibfnamefont{B.~D.} \bibnamefont{{Wandelt}}},
  \bibnamefont{and} \bibinfo{author}{\bibfnamefont{C.~M.}
  \bibnamefont{{Hirata}}}, \bibinfo{journal}{\prd}
  \textbf{\bibinfo{volume}{92}}, \bibinfo{eid}{083531}
  (\bibinfo{year}{2015}{\natexlab{a}}), \eprint{1503.07690}.

\bibitem[{\citenamefont{{Sahl{\'e}n} et~al.}(2016)\citenamefont{{Sahl{\'e}n},
  {Zubeld{\'{i}}a}, and {Silk}}}]{Sahlen2016}
\bibinfo{author}{\bibfnamefont{M.}~\bibnamefont{{Sahl{\'e}n}}},
  \bibinfo{author}{\bibfnamefont{{\'{I}}.}~\bibnamefont{{Zubeld{\'{i}}a}}},
  \bibnamefont{and} \bibinfo{author}{\bibfnamefont{J.}~\bibnamefont{{Silk}}},
  \bibinfo{journal}{\apjl} \textbf{\bibinfo{volume}{820}}, \bibinfo{eid}{L7}
  (\bibinfo{year}{2016}), \eprint{1511.04075}.

\bibitem[{\citenamefont{{Zivick} et~al.}(2015)\citenamefont{{Zivick}, {Sutter},
  {Wandelt}, {Li}, and {Lam}}}]{Zivick2015}
\bibinfo{author}{\bibfnamefont{P.}~\bibnamefont{{Zivick}}},
  \bibinfo{author}{\bibfnamefont{P.~M.} \bibnamefont{{Sutter}}},
  \bibinfo{author}{\bibfnamefont{B.~D.} \bibnamefont{{Wandelt}}},
  \bibinfo{author}{\bibfnamefont{B.}~\bibnamefont{{Li}}}, \bibnamefont{and}
  \bibinfo{author}{\bibfnamefont{T.~Y.} \bibnamefont{{Lam}}},
  \bibinfo{journal}{\mnras} \textbf{\bibinfo{volume}{451}},
  \bibinfo{pages}{4215} (\bibinfo{year}{2015}), \eprint{1411.5694}.

\bibitem[{\citenamefont{{Barreira} et~al.}(2015)\citenamefont{{Barreira},
  {Cautun}, {Li}, {Baugh}, and {Pascoli}}}]{Barreira2015}
\bibinfo{author}{\bibfnamefont{A.}~\bibnamefont{{Barreira}}},
  \bibinfo{author}{\bibfnamefont{M.}~\bibnamefont{{Cautun}}},
  \bibinfo{author}{\bibfnamefont{B.}~\bibnamefont{{Li}}},
  \bibinfo{author}{\bibfnamefont{C.~M.} \bibnamefont{{Baugh}}},
  \bibnamefont{and}
  \bibinfo{author}{\bibfnamefont{S.}~\bibnamefont{{Pascoli}}},
  \bibinfo{journal}{\jcap} \textbf{\bibinfo{volume}{8}}, \bibinfo{eid}{028}
  (\bibinfo{year}{2015}), \eprint{1505.05809}.

\bibitem[{\citenamefont{{Massara} et~al.}(2015)\citenamefont{{Massara},
  {Villaescusa-Navarro}, {Viel}, and {Sutter}}}]{Massara2015}
\bibinfo{author}{\bibfnamefont{E.}~\bibnamefont{{Massara}}},
  \bibinfo{author}{\bibfnamefont{F.}~\bibnamefont{{Villaescusa-Navarro}}},
  \bibinfo{author}{\bibfnamefont{M.}~\bibnamefont{{Viel}}}, \bibnamefont{and}
  \bibinfo{author}{\bibfnamefont{P.~M.} \bibnamefont{{Sutter}}},
  \bibinfo{journal}{\jcap} \textbf{\bibinfo{volume}{11}}, \bibinfo{eid}{018}
  (\bibinfo{year}{2015}), \eprint{1506.03088}.

\bibitem[{\citenamefont{{Dawson} et~al.}(2013)\citenamefont{{Dawson},
  {Schlegel}, {Ahn}, {Anderson}, {Aubourg}, {Bailey}, {Barkhouser}, {Bautista},
  {Beifiori}, {Berlind} et~al.}}]{Dawson2013}
\bibinfo{author}{\bibfnamefont{K.~S.} \bibnamefont{{Dawson}}},
  \bibinfo{author}{\bibfnamefont{D.~J.} \bibnamefont{{Schlegel}}},
  \bibinfo{author}{\bibfnamefont{C.~P.} \bibnamefont{{Ahn}}},
  \bibinfo{author}{\bibfnamefont{S.~F.} \bibnamefont{{Anderson}}},
  \bibinfo{author}{\bibfnamefont{{\'E}.}~\bibnamefont{{Aubourg}}},
  \bibinfo{author}{\bibfnamefont{S.}~\bibnamefont{{Bailey}}},
  \bibinfo{author}{\bibfnamefont{R.~H.} \bibnamefont{{Barkhouser}}},
  \bibinfo{author}{\bibfnamefont{J.~E.} \bibnamefont{{Bautista}}},
  \bibinfo{author}{\bibfnamefont{A.}~\bibnamefont{{Beifiori}}},
  \bibinfo{author}{\bibfnamefont{A.~A.} \bibnamefont{{Berlind}}},
  \bibnamefont{et~al.}, \bibinfo{journal}{\aj} \textbf{\bibinfo{volume}{145}},
  \bibinfo{eid}{10} (\bibinfo{year}{2013}), \eprint{1208.0022}.

\bibitem[{\citenamefont{{Eisenstein} et~al.}(2011)\citenamefont{{Eisenstein},
  {Weinberg}, {Agol}, {Aihara}, {Allende Prieto}, {Anderson}, {Arns},
  {Aubourg}, {Bailey}, {Balbinot} et~al.}}]{Eisenstein2011}
\bibinfo{author}{\bibfnamefont{D.~J.} \bibnamefont{{Eisenstein}}},
  \bibinfo{author}{\bibfnamefont{D.~H.} \bibnamefont{{Weinberg}}},
  \bibinfo{author}{\bibfnamefont{E.}~\bibnamefont{{Agol}}},
  \bibinfo{author}{\bibfnamefont{H.}~\bibnamefont{{Aihara}}},
  \bibinfo{author}{\bibfnamefont{C.}~\bibnamefont{{Allende Prieto}}},
  \bibinfo{author}{\bibfnamefont{S.~F.} \bibnamefont{{Anderson}}},
  \bibinfo{author}{\bibfnamefont{J.~A.} \bibnamefont{{Arns}}},
  \bibinfo{author}{\bibfnamefont{{\'E}.}~\bibnamefont{{Aubourg}}},
  \bibinfo{author}{\bibfnamefont{S.}~\bibnamefont{{Bailey}}},
  \bibinfo{author}{\bibfnamefont{E.}~\bibnamefont{{Balbinot}}},
  \bibnamefont{et~al.}, \bibinfo{journal}{\aj} \textbf{\bibinfo{volume}{142}},
  \bibinfo{eid}{72} (\bibinfo{year}{2011}), \eprint{1101.1529}.

\bibitem[{\citenamefont{{Alam} et~al.}(2015)\citenamefont{{Alam}, {Albareti},
  {Allende Prieto}, {Anders}, {Anderson}, {Anderton}, {Andrews}, {Armengaud},
  {Aubourg}, {Bailey} et~al.}}]{Alam2015a}
\bibinfo{author}{\bibfnamefont{S.}~\bibnamefont{{Alam}}},
  \bibinfo{author}{\bibfnamefont{F.~D.} \bibnamefont{{Albareti}}},
  \bibinfo{author}{\bibfnamefont{C.}~\bibnamefont{{Allende Prieto}}},
  \bibinfo{author}{\bibfnamefont{F.}~\bibnamefont{{Anders}}},
  \bibinfo{author}{\bibfnamefont{S.~F.} \bibnamefont{{Anderson}}},
  \bibinfo{author}{\bibfnamefont{T.}~\bibnamefont{{Anderton}}},
  \bibinfo{author}{\bibfnamefont{B.~H.} \bibnamefont{{Andrews}}},
  \bibinfo{author}{\bibfnamefont{E.}~\bibnamefont{{Armengaud}}},
  \bibinfo{author}{\bibfnamefont{{\'E}.}~\bibnamefont{{Aubourg}}},
  \bibinfo{author}{\bibfnamefont{S.}~\bibnamefont{{Bailey}}},
  \bibnamefont{et~al.}, \bibinfo{journal}{\apjs}
  \textbf{\bibinfo{volume}{219}}, \bibinfo{eid}{12} (\bibinfo{year}{2015}),
  \eprint{1501.00963}.

\bibitem[{\citenamefont{{Anderson} et~al.}(2014)\citenamefont{{Anderson},
  {Aubourg}, {Bailey}, {Beutler}, {Bhardwaj}, {Blanton}, {Bolton}, {Brinkmann},
  {Brownstein}, {Burden} et~al.}}]{Anderson2014}
\bibinfo{author}{\bibfnamefont{L.}~\bibnamefont{{Anderson}}},
  \bibinfo{author}{\bibfnamefont{{\'E}.}~\bibnamefont{{Aubourg}}},
  \bibinfo{author}{\bibfnamefont{S.}~\bibnamefont{{Bailey}}},
  \bibinfo{author}{\bibfnamefont{F.}~\bibnamefont{{Beutler}}},
  \bibinfo{author}{\bibfnamefont{V.}~\bibnamefont{{Bhardwaj}}},
  \bibinfo{author}{\bibfnamefont{M.}~\bibnamefont{{Blanton}}},
  \bibinfo{author}{\bibfnamefont{A.~S.} \bibnamefont{{Bolton}}},
  \bibinfo{author}{\bibfnamefont{J.}~\bibnamefont{{Brinkmann}}},
  \bibinfo{author}{\bibfnamefont{J.~R.} \bibnamefont{{Brownstein}}},
  \bibinfo{author}{\bibfnamefont{A.}~\bibnamefont{{Burden}}},
  \bibnamefont{et~al.}, \bibinfo{journal}{\mnras}
  \textbf{\bibinfo{volume}{441}}, \bibinfo{pages}{24} (\bibinfo{year}{2014}),
  \eprint{1312.4877}.

\bibitem[{\citenamefont{{Sutter} et~al.}(2015)\citenamefont{{Sutter}, {Lavaux},
  {Hamaus}, {Pisani}, {Wandelt}, {Warren}, {Villaescusa-Navarro}, {Zivick},
  {Mao}, and {Thompson}}}]{Sutter2014d}
\bibinfo{author}{\bibfnamefont{P.~M.} \bibnamefont{{Sutter}}},
  \bibinfo{author}{\bibfnamefont{G.}~\bibnamefont{{Lavaux}}},
  \bibinfo{author}{\bibfnamefont{N.}~\bibnamefont{{Hamaus}}},
  \bibinfo{author}{\bibfnamefont{A.}~\bibnamefont{{Pisani}}},
  \bibinfo{author}{\bibfnamefont{B.~D.} \bibnamefont{{Wandelt}}},
  \bibinfo{author}{\bibfnamefont{M.}~\bibnamefont{{Warren}}},
  \bibinfo{author}{\bibfnamefont{F.}~\bibnamefont{{Villaescusa-Navarro}}},
  \bibinfo{author}{\bibfnamefont{P.}~\bibnamefont{{Zivick}}},
  \bibinfo{author}{\bibfnamefont{Q.}~\bibnamefont{{Mao}}}, \bibnamefont{and}
  \bibinfo{author}{\bibfnamefont{B.~B.} \bibnamefont{{Thompson}}},
  \bibinfo{journal}{\ac} \textbf{\bibinfo{volume}{9}}, \bibinfo{pages}{1}
  (\bibinfo{year}{2015}), \eprint{1406.1191}.

\bibitem[{\citenamefont{{Neyrinck}}(2008)}]{Neyrinck2008}
\bibinfo{author}{\bibfnamefont{M.~C.} \bibnamefont{{Neyrinck}}},
  \bibinfo{journal}{\mnras} \textbf{\bibinfo{volume}{386}},
  \bibinfo{pages}{2101} (\bibinfo{year}{2008}), \eprint{0712.3049}.

\bibitem[{\citenamefont{{Sutter}
  et~al.}(2012{\natexlab{b}})\citenamefont{{Sutter}, {Lavaux}, {Wandelt}, and
  {Weinberg}}}]{Sutter2012a}
\bibinfo{author}{\bibfnamefont{P.~M.} \bibnamefont{{Sutter}}},
  \bibinfo{author}{\bibfnamefont{G.}~\bibnamefont{{Lavaux}}},
  \bibinfo{author}{\bibfnamefont{B.~D.} \bibnamefont{{Wandelt}}},
  \bibnamefont{and} \bibinfo{author}{\bibfnamefont{D.~H.}
  \bibnamefont{{Weinberg}}}, \bibinfo{journal}{\apj}
  \textbf{\bibinfo{volume}{761}}, \bibinfo{eid}{44}
  (\bibinfo{year}{2012}{\natexlab{b}}), \eprint{1207.2524}.

\bibitem[{\citenamefont{{Sutter}
  et~al.}(2014{\natexlab{b}})\citenamefont{{Sutter}, {Lavaux}, {Wandelt},
  {Weinberg}, {Warren}, and {Pisani}}}]{Sutter2013b}
\bibinfo{author}{\bibfnamefont{P.~M.} \bibnamefont{{Sutter}}},
  \bibinfo{author}{\bibfnamefont{G.}~\bibnamefont{{Lavaux}}},
  \bibinfo{author}{\bibfnamefont{B.~D.} \bibnamefont{{Wandelt}}},
  \bibinfo{author}{\bibfnamefont{D.~H.} \bibnamefont{{Weinberg}}},
  \bibinfo{author}{\bibfnamefont{M.~S.} \bibnamefont{{Warren}}},
  \bibnamefont{and} \bibinfo{author}{\bibfnamefont{A.}~\bibnamefont{{Pisani}}},
  \bibinfo{journal}{\mnras} \textbf{\bibinfo{volume}{442}},
  \bibinfo{pages}{3127} (\bibinfo{year}{2014}{\natexlab{b}}),
  \eprint{1310.7155}.

\bibitem[{\citenamefont{{Pisani et al.}}()}]{Pisani2016}
\bibinfo{author}{\bibfnamefont{A.}~\bibnamefont{{Pisani et al.}}}, \eprint{(to
  be published).}

\bibitem[{\citenamefont{{Patil} et~al.}(2010)\citenamefont{{Patil}, {Huard},
  and {Fonnesbeck}}}]{Patil2010}
\bibinfo{author}{\bibfnamefont{A.}~\bibnamefont{{Patil}}},
  \bibinfo{author}{\bibfnamefont{D.}~\bibnamefont{{Huard}}}, \bibnamefont{and}
  \bibinfo{author}{\bibfnamefont{C.~J.} \bibnamefont{{Fonnesbeck}}},
  \bibinfo{journal}{\jsts} \textbf{\bibinfo{volume}{35}}, \bibinfo{pages}{1}
  (\bibinfo{year}{2010}).

\bibitem[{\citenamefont{{Sutter}
  et~al.}(2014{\natexlab{c}})\citenamefont{{Sutter}, {Lavaux}, {Hamaus},
  {Wandelt}, {Weinberg}, and {Warren}}}]{Sutter2014a}
\bibinfo{author}{\bibfnamefont{P.~M.} \bibnamefont{{Sutter}}},
  \bibinfo{author}{\bibfnamefont{G.}~\bibnamefont{{Lavaux}}},
  \bibinfo{author}{\bibfnamefont{N.}~\bibnamefont{{Hamaus}}},
  \bibinfo{author}{\bibfnamefont{B.~D.} \bibnamefont{{Wandelt}}},
  \bibinfo{author}{\bibfnamefont{D.~H.} \bibnamefont{{Weinberg}}},
  \bibnamefont{and} \bibinfo{author}{\bibfnamefont{M.~S.}
  \bibnamefont{{Warren}}}, \bibinfo{journal}{\mnras}
  \textbf{\bibinfo{volume}{442}}, \bibinfo{pages}{462}
  (\bibinfo{year}{2014}{\natexlab{c}}), \eprint{1309.5087}.

\bibitem[{\citenamefont{{Paz} and {S{\'a}nchez}}(2015)}]{Paz2015}
\bibinfo{author}{\bibfnamefont{D.~J.} \bibnamefont{{Paz}}} \bibnamefont{and}
  \bibinfo{author}{\bibfnamefont{A.~G.} \bibnamefont{{S{\'a}nchez}}},
  \bibinfo{journal}{\mnras} \textbf{\bibinfo{volume}{454}},
  \bibinfo{pages}{4326} (\bibinfo{year}{2015}), \eprint{1508.03162}.

\bibitem[{\citenamefont{{Pisani}
  et~al.}(2015{\natexlab{b}})\citenamefont{{Pisani}, {Sutter}, and
  {Wandelt}}}]{Pisani2015b}
\bibinfo{author}{\bibfnamefont{A.}~\bibnamefont{{Pisani}}},
  \bibinfo{author}{\bibfnamefont{P.~M.} \bibnamefont{{Sutter}}},
  \bibnamefont{and} \bibinfo{author}{\bibfnamefont{B.~D.}
  \bibnamefont{{Wandelt}}}, \bibinfo{journal}{ArXiv e-prints}
  (\bibinfo{year}{2015}{\natexlab{b}}), \eprint{1506.07982}.

\bibitem[{\citenamefont{{Manera} et~al.}(2013)\citenamefont{{Manera},
  {Scoccimarro}, {Percival}, {Samushia}, {McBride}, {Ross}, {Sheth}, {White},
  {Reid}, {S{\'a}nchez} et~al.}}]{Manera2013}
\bibinfo{author}{\bibfnamefont{M.}~\bibnamefont{{Manera}}},
  \bibinfo{author}{\bibfnamefont{R.}~\bibnamefont{{Scoccimarro}}},
  \bibinfo{author}{\bibfnamefont{W.~J.} \bibnamefont{{Percival}}},
  \bibinfo{author}{\bibfnamefont{L.}~\bibnamefont{{Samushia}}},
  \bibinfo{author}{\bibfnamefont{C.~K.} \bibnamefont{{McBride}}},
  \bibinfo{author}{\bibfnamefont{A.~J.} \bibnamefont{{Ross}}},
  \bibinfo{author}{\bibfnamefont{R.~K.} \bibnamefont{{Sheth}}},
  \bibinfo{author}{\bibfnamefont{M.}~\bibnamefont{{White}}},
  \bibinfo{author}{\bibfnamefont{B.~A.} \bibnamefont{{Reid}}},
  \bibinfo{author}{\bibfnamefont{A.~G.} \bibnamefont{{S{\'a}nchez}}},
  \bibnamefont{et~al.}, \bibinfo{journal}{\mnras}
  \textbf{\bibinfo{volume}{428}}, \bibinfo{pages}{1036} (\bibinfo{year}{2013}),
  \eprint{1203.6609}.

\bibitem[{\citenamefont{{P.A.R. Ade et al. (\textsc{Planck}
  Collaboration)}}(2014)}]{Planck2013}
\bibinfo{author}{\bibnamefont{{P.A.R. Ade et al. (\textsc{Planck}
  Collaboration)}}}, \bibinfo{journal}{\aap} \textbf{\bibinfo{volume}{571}},
  \bibinfo{eid}{A16} (\bibinfo{year}{2014}), \eprint{1303.5076}.

\bibitem[{\citenamefont{{Mao} et~al.}(2016)\citenamefont{{Mao}, {Berlind},
  {Scherrer}, {Neyrinck}, {Scoccimarro}, {Tinker}, and {McBride}}}]{Mao2016}
\bibinfo{author}{\bibfnamefont{Q.}~\bibnamefont{{Mao}}},
  \bibinfo{author}{\bibfnamefont{A.~A.} \bibnamefont{{Berlind}}},
  \bibinfo{author}{\bibfnamefont{R.~J.} \bibnamefont{{Scherrer}}},
  \bibinfo{author}{\bibfnamefont{M.~C.} \bibnamefont{{Neyrinck}}},
  \bibinfo{author}{\bibfnamefont{R.}~\bibnamefont{{Scoccimarro}}},
  \bibinfo{author}{\bibfnamefont{J.~L.} \bibnamefont{{Tinker}}},
  \bibnamefont{and} \bibinfo{author}{\bibfnamefont{C.~K.}
  \bibnamefont{{McBride}}}, \bibinfo{journal}{ArXiv e-prints}
  (\bibinfo{year}{2016}), \eprint{1602.06306}.

\end{thebibliography}
\bibliographystyle{apsrev.bst}



{\onecolumngrid
\clearpage
\begin{center}
\textbf{\large Supplemental Material}
\end{center}
\vspace{0.2cm}}

\twocolumngrid

\noindent\textit{Data.}---For our analysis we use public data from the Baryon Oscillation Spectroscopic Survey (BOSS)~\cite{Dawson2013} of the SDSS-III~\cite{Eisenstein2011}, more precisely the CMASS galaxy sample from Data Release 11 (DR11)~\cite{Alam2015a}. This sample is spread over a redshift range of $0.43<z<0.7$ with a median of $\bar{z}=0.57$ inside a total volume of about $3.5h^{-3}\mathrm{Gpc}^3$, with a peak number density of roughly $4\times10^{-4}h^3\mathrm{Mpc}^{-3}$ and a linear bias parameter of $b\simeq1.87$~\cite{Anderson2014}. We use the \textsc{vide} software~\cite{Sutter2014d} to generate void catalogs, it is based on the implementation of a \emph{watershed} algorithm provided by the code \textsc{zobov}~\cite{Neyrinck2008}. A detailed description of this procedure can be found in Refs.~\cite{Sutter2012a,Sutter2013b} for DR7 and DR9, and in Ref.~\cite{Pisani2016} for DR11. Whenever we apply coordinate transformations via Eq.~(\ref{coords}), we assume the following fiducial cosmological parameters: $\Omega_\matter=0.27$, $\Omega_\Lambda=0.73$, $\Omega_k=0$, and $h=0.70$. The resulting void catalogs provide us with the sky-coordinates and redshifts of each void's volume-weighted barycenter, as well as its effective radius $r_\void$ and volume $V_\void$, among many other properties. Besides insisting on $r_\void$ to be at least as large as the mean galaxy separation in the sample to avoid Poisson contamination, we apply no further post-processing cuts.

This results in a catalog of $3457$ voids with effective radius range $16.2\hMpc<r_\void<98.0\hMpc$. We split the full range of void radii into $8$ adjacent bins such that every bin contains roughly the same number of voids. In each bin, all void centers and their surrounding galaxies that are within a distance of $3r_\void$ are aligned with the line-of-sight direction and stacked. We only consider voids that do not intersect with any survey boundaries to avoid the inclusion of incomplete voids and unobserved regions in the stacks. Each stack is then histogrammed in two directions: the void-centric distances along and perpendicular to the line of sight, $r_\parallel$ and $r_\perp$, which yields an estimator of the void-galaxy cross-correlation function in redshift space. For more details on this procedure we refer the reader to Ref.~\cite{Hamaus2015}.\newline

\noindent\textit{Analysis.}---For the comparison of our model from Eq.~(\ref{GSM}) with the observational data in Fig.~\ref{stacks}, we employ a MCMC technique using a \emph{Metropolis-Hastings} sampler, implemented in the software package PyMC~\cite{Patil2010}. Assuming Gaussian statistics, the likelihood can be expressed as
\begin{equation*}
 \mathcal{L}(\hat{\xi}_{\void\gal}|\boldsymbol{\theta}) \propto \exp\left[-\frac{1}{2}(\hat{\xi}_{\void\gal}-\xi_{\void\gal})^\intercal\mathbf{C}^{-1}(\hat{\xi}_{\void\gal}-\xi_{\void\gal})\right], \label{likelihood}
\end{equation*}
where a $\hat{\xi}_{\void\gal}$ denotes the measured void-galaxy cross-correlation function, $\mathbf{C}$ its covariance matrix, and $\boldsymbol{\theta} = (\rs,\dc,\alpha,\beta,\sigma_v,f/b,\Omega_\matter)$ the parameter vector of our model. Furthermore, we assume $\Omega_k=0$ and set $H_0=100\,\mathrm{km}\,\mathrm{s}^{-1}h\mathrm{Mpc}^{-1}$, while expressing all distances in units of $\hMpc$. The remaining cosmological parameters $n_s$ and $\sigma_8$ do not appear explicitly here, but their influence is captured by the void density-profile parameters $(\rs,\dc,\alpha,\beta)$. For example, $\sigma_8$ determines the amplitude of density fluctuations and is therefore degenerate with $\dc$ and $b$.

\begin{figure}[p]
\vspace{2cm}
\resizebox{\hsize}{!}{\includegraphics{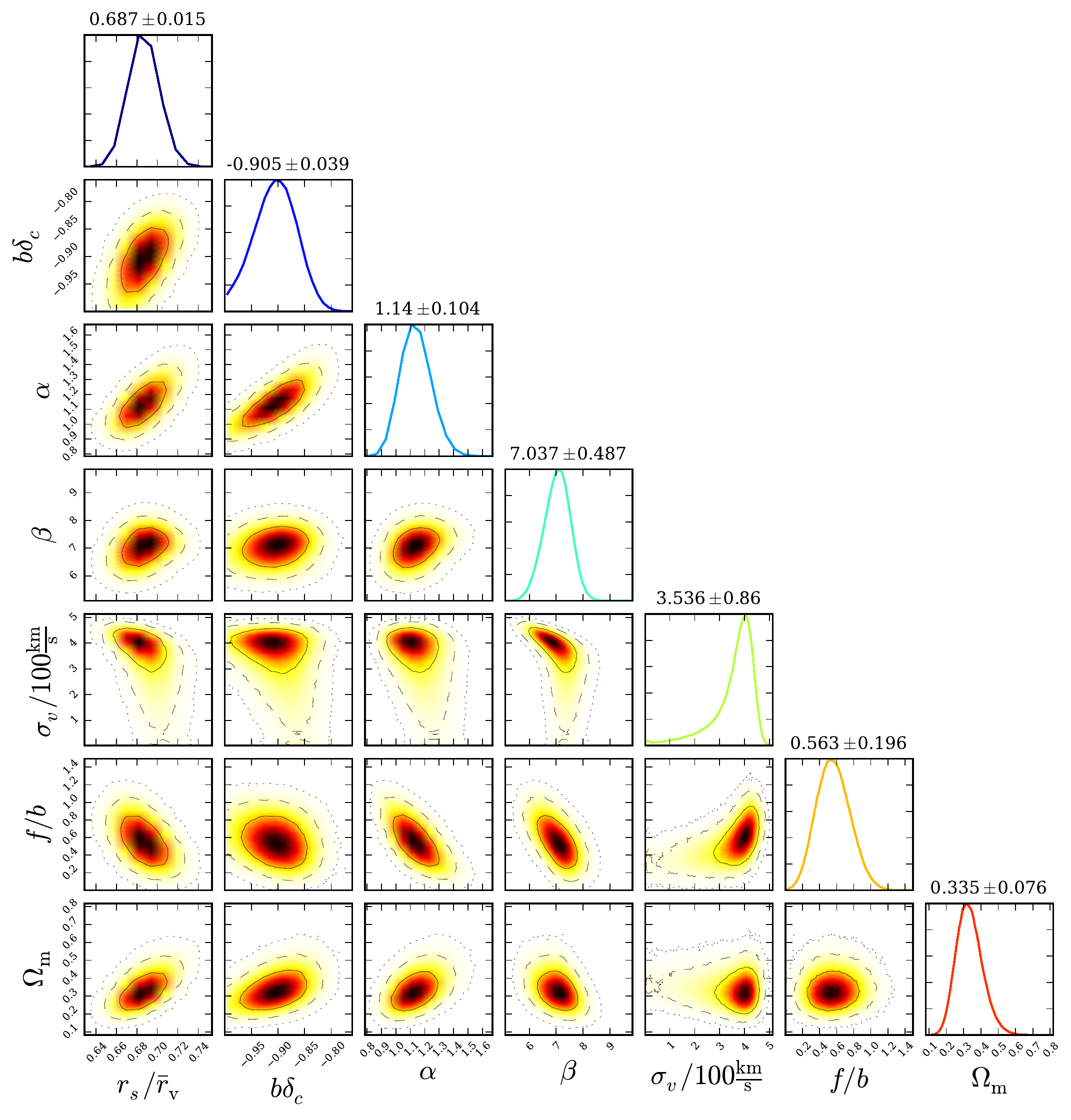}}
\caption{Complete posterior distribution of all model parameters for a void stack of size range $r_\void=49.4\hMpc - 57.2\hMpc$.}
\label{triangle}
\end{figure}

Our constraints on the void density-profile parameters are consistent with what has been presented in the mock analysis of Ref.~\cite{Hamaus2015}. Fig.~\ref{triangle} presents the complete posterior distribution of all relevant parameters obtained from one of our void stacks with effective radius range $r_\void=49.4\hMpc - 57.2\hMpc$. In general, we find a very good agreement with dedicated studies on the void density-profile~\cite{Hamaus2014b,Sutter2014a}. We also marginalize over a constant velocity dispersion $\sigma_v$, a simplification that was shown not to influence any of the other parameter constraints of the model~\cite{Hamaus2015}. The covariance matrix is estimated via Jackknife resampling of the voids in each stack, and inverted using the \emph{tapering} technique~\cite{Paz2015} (details in Ref.~\cite{Hamaus2015}). Imposing uniform prior distributions with sufficiently wide ranges for our model parameters, we estimate their posterior distribution by running MCMC chains of $\mathcal{O}(10^6)$ samples. For every void stack, we evaluate a best-fit model from the parameter set in the chain that yields the highest likelihood, as depicted in Fig.~\ref{stacks}.\newline

\noindent\textit{Discussion.}---In order to check for systematics in our measurement, we have conducted a number of tests. A potential problem can arise from the inclusion of voids whose effective radius is close to the mean galaxy separation in the survey, both because of Poisson contamination, a redshift-dependent average galaxy density, and the effects of nonlinear RSDs (FoG)~\cite{Pisani2015b}. We repeated our analysis after removing all voids with effective radii below twice the average galaxy separation, i.e. $r_\void<30\hMpc$. However, the final cosmological constraints are hardly affected by this stricter size-cut, since the information content from the smaller voids is relatively weak anyway, as can be seen in Fig.~\ref{pdfs}.

\begin{figure}[p]
\vspace{2cm}
\resizebox{\hsize}{!}{\includegraphics{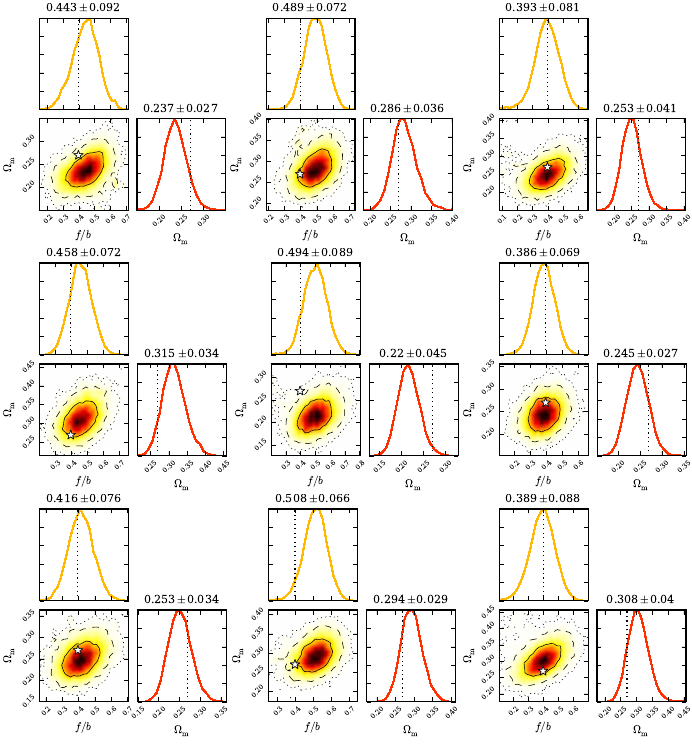}}
\caption{Same as Fig.~\ref{pdfall}, but for $9$ different bootstrap realizations of the original void catalog.}
\label{bootstraps}
\end{figure}

As a further test concerning the significance of our results we performed bootstrap resampling of our original void catalog, i.e., randomly selecting the same number of voids with replacement. We generated $9$ such bootstraps and repeated the inference process from all voids for each of the bootstrap realizations, the results are presented in Fig.~\ref{bootstraps}. The solid, dashed, and dotted contour levels correspond to $68.3\%$, $95.5\%$, and $99.7\%$ confidence regions. Among all bootstraps, the fiducial parameters agree to $6/9$ with the solid contour, to $8/9$ with the dashed contour, and only $1/9$ marginally lies outside the latter, but still within the dotted confidence level. Thus, the statistical fluctuations in the final constraints are entirely consistent with the expectation, providing further confidence in our constraints.

\begin{figure}[p]
\vspace{2cm}
\resizebox{\hsize}{!}{\includegraphics{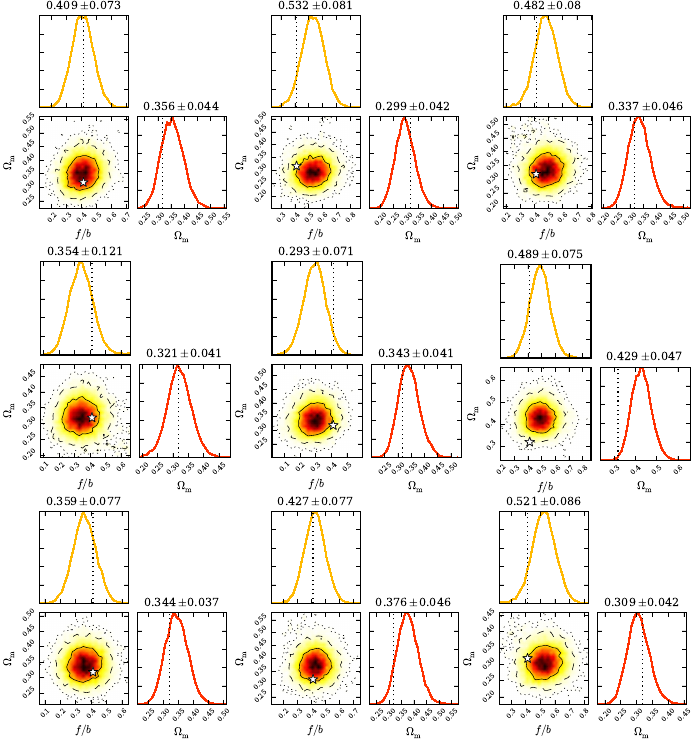}}
\caption{Same as Fig.~\ref{pdfall}, but for $9$ different bootstrap realizations of a mock void catalog with $\Omega_\matter=0.32$ and $f/b=0.41$.}
\label{mocks}
\end{figure}

We also cross-checked our measurement with the help of a simulated mock-galaxy catalog whose properties closely resemble the observed galaxy sample from this analysis. In particular, we used a common \emph{halo occupation distribution} (HOD) model to populate dark matter halos from an $N$-body simulation with central and satellite galaxies, the HOD has been calibrated to the CMASS galaxies from the SDSS-III DR9~\cite{Manera2013} resulting in a mean galaxy density of $3\times10^{-4}h^3\mathrm{Mpc}^{-3}$ and a bias parameter of $b\simeq1.84$. The simulation covers a cubic box of volume $1h^{-3}\mathrm{Gpc}^3$ at redshift $z=0.5$ and adopts a Planck 2013 cosmology~\cite{Planck2013} with $\Omega_\matter=0.32$, $\Omega_\Lambda=0.68$, $\Omega_k=0$, and $h=0.68$ (see Ref.~\cite{Hamaus2015} for more details). From this mock-galaxy sample we identify $2559$ voids and repeat our analysis with $9$ different bootstraps from this catalog, the results of which are shown in Fig.~\ref{mocks}. The final parameter constraints and statistics are fully in line with the measurement from the observed data above. Again we find $6$ out of $9$ bootstraps to agree with the input cosmology within the $68.3\%$ confidence region, $8$ out of $9$ within the $95.5\%$ one, leaving one bootstrap to agree only within $99.7\%$ slightly outside the $95.5\%$ contour level.

\begin{figure}[h]
\vspace{1cm}
\resizebox{\hsize}{!}{\includegraphics{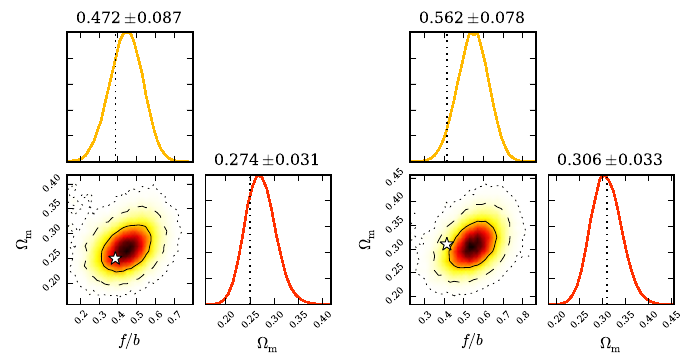}}
\caption{Same as Fig.~\ref{pdfall}, but assuming two different fiducial cosmologies with $\Omega_\matter=0.25$ (left) and $\Omega_\matter=0.31$ (right) in the analysis. The corresponding fiducial values for the growth rate are $f/b = 0.39$ (left) and $f/b=0.41$ (right), respectively.}
\label{cosmotest}
\end{figure}

Finally, we tested the robustness of our results when varying the assumed fiducial cosmology. In a recent study involving cosmic voids from the SDSS DR12 BOSS data it has been pointed out that the expected sensitivity of the AP test can be significantly degraded due to imperfect void identification in sparsely sampled galaxies~\cite{Mao2016}. We therefore repeated our entire analysis with two new fiducial values for $\Omega_\matter$ that differ from the obtained value of $\Omega_\matter = 0.281\pm0.031$ by $\pm1\sigma$, namely $\Omega_\matter = 0.25$ and $\Omega_\matter = 0.31$. The resulting posterior distributions for $\Omega_\matter$ and $f/b$ are shown in Fig.~\ref{cosmotest} for each case. We do observe some dependence on the fiducial cosmology, especially towards higher values of $\Omega_\matter$, which somewhat degrades our constraints on that parameter. Comparing the spread of $2\sigma$ in its fiducial values to the resulting spread of about $1\sigma$ yields an estimate of $0.5\sigma$ for the additional systematic error on $\Omega_\matter$. The fixed point in the mapping between its fiducial and measured values yields $\Omega_\matter\simeq0.30$, which is still within $\frac{2}{3}\sigma$ of the originally quoted value. Nonetheless, it is interesting to note that a tension with the fiducial value of $f/b$ is generated at the same time. The combined constraints on $\Omega_\matter$ and $f/b$ are thus much less affected by this issue, if GR is assumed to be the correct theory of gravity. We would like to point out, however, that to the best of our knowledge the standard procedure in joint AP/RSD analyses based on galaxy two-point statistics does not account for the mentioned issue either. This would require a recomputation of the estimators, optimal weights, mocks, covariances, etc. for each cosmology. In that sense the analysis presented by Ref.~\cite{Mao2016} goes beyond the traditional methodology.

\onecolumngrid

\end{document}